\documentclass{aa} % for the letters

\usepackage{graphicx,subfigure}
\usepackage{pstricks}
\usepackage{ulem}

  \usepackage{natbib}
  \bibpunct{[}{]}{;}{a}{}{,}

\graphicspath{{figures/}}

\usepackage{hyperref}
\hypersetup{
    bookmarks=true,         % show bookmarks bar?
    unicode=false,          % non-Latin characters in Acrobat's bookmarks
    pdftoolbar=true,        % show Acrobat's toolbar?
    pdfmenubar=true,        % show Acrobat's menu?
    pdffitwindow=false,     % page fit to window when opened
    pdftitle={Physical properties of (21) Lutetia},
    pdfauthor={Jack Drummond},
    pdfsubject={Planetary Science},
    pdfkeywords={},         % list of keywords
    pdfnewwindow=true,      % links in new window
    colorlinks=true,        % false: boxed links; true: colored links
    linkcolor=gray,         % color of internal links
    citecolor=blue,         % color of links to bibliography
    filecolor=gray,         % color of file links
    urlcolor=gray           % color of external links
}

\usepackage{lscape,epsf}

\def\"{^{\prime \prime}}
\def\deg{^{\circ}}

\begin{document}
  \title{The triaxial ellipsoid dimensions, rotational pole, and bulk density of ESA Rosetta target asteroid (21) Lutetia
    \thanks{Based on observations
      collected at the W. M. Keck Observatory and the
      European Southern Observatory
      Very Large Telescope (program ID:
      \href{http://archive.eso.org/wdb/wdb/eso/eso_archive_main/query?prog_id=079.C-0493\%28A\%29\&max_rows_returned=1000}{079.C-0493},
      PI: E.~Dotto). The W. M. Keck Observatory is operated as a scientific partnership among the California Institute of Technology, the University of California, and the National Aeronautics and Space Administration. The Observatory
      was made possible by the generous financial support of the W. M. Keck Foundation.}}

  \author{%
    Jack D.~Drummond\inst{1}
    \and A.~Conrad\inst{2}
    \and W.~J.~Merline\inst{3}
    \and B.~Carry\inst{4,5}
    \and C.~R.~Chapman\inst{3}
    \and H.~A.~Weaver\inst{6}
    \and P.~M.~Tamblyn\inst{3}
    \and J.~C.~Christou\inst{7}
    \and C.~Dumas\inst{8}
%    \and E.~Dotto\inst{9}
%    \and D.~Perna\inst{4,9,10}
%    \and S.~Fornasier\inst{4,5}
    %\and M.~J.~Mutchler\inst{10}
    }

  \offprints{Jack.Drummond@Kirtland.af.mil}

  \institute{%
    Starfire Optical Range, Directed Energy Directorate, Air Force
    Research Laboratory, 3550 Aberdeen Av SE, Kirtland AFB, New Mexico
    87117-5776, USA
    \and W.M. Keck Observatory, 65-1120 Mamalahoa Highway, Kamuela, HI, 96743, USA
    \and Southwest Research Institute, 1050 Walnut Street, Suite 300, Boulder, CO 80302, USA
    \and LESIA, Observatoire de Paris, 5 place Jules Janssen, 92190 MEUDON, France
    \and Universit\'e Paris 7 Denis-Diderot, 5 rue Thomas Mann, 75205 PARIS CEDEX, France
    \and Johns Hopkins University Applied Physics Laboratory, Laurel, MD 20723-6099, USA
    \and Gemini Observatory, 670 N. A'ohoku Place, Hilo, Hawaii, 96720, USA
    \and ESO, Alonso de C´ordova 3107, Vitacura, Casilla 19001, Santiago de Chile, Chile
%    \and INAF, Osservatorio Astronomico di Roma, via Frascati 33, 00040
%    Monteporzio Catone (Roma), Italy
%    \and Dipartimento di Fisica, Universit\`a di Roma Tor Vergata, Via %della Ricerca Scientifica 1, 00133 Roma, Italy
    %\and tbd
}

  \date{Received September 15, 1996; accepted March 16, 1997}

% \abstract{}{}{}{}{}
% 5 {} token are mandatory

  \abstract
      {Asteroid (21) Lutetia is the target of the ESA Rosetta
        mission flyby in 2010 July.}
      {We seek the best size estimates of the asteroid, the
        direction of its spin axis, and its bulk density,
        assuming its
        shape is well described by a smooth featureless triaxial ellipsoid, and to evaluate the deviations from this assumption.}
      {We derive these quantities from the outlines of the asteroid in
%        307 adaptive optics images,
        307 images of its resolved apparent disk obtained with
        adaptive optics (AO) at Keck II and VLT,
        and combine these with recent mass
        determinations to estimate a bulk density.}
      {Our best triaxial ellipsoid diameters for Lutetia, based on
        our AO images alone, are
        $a\times b\times c = 132\times 101\times 93$ km, with uncertainties of
        $4\times3\times13$ km including estimated systematics, with a
        rotational pole within 5$\deg$ of
        ECJ2000 [$\lambda\ \beta$] = [$45\deg -7\deg$],
        or EQJ2000 [RA DEC] = [$44\deg\ +9\deg$].
        The AO model fit itself has internal precisions of $1\times1\times8$ km,
        but it is evident, both from this model derived from limited
        viewing aspects and the radius vector model given in a companion paper,
        that Lutetia has significant departures from an idealized ellipsoid. In
        particular, the long axis may be overestimated from the AO images alone
        by about 10 km. Therefore, we combine the best aspects of the radius
        vector and ellipsoid model into a hybrid ellipsoid model, as our final
        result, of $124\pm5\times101\pm4\times93\pm13$ km that can be used
        to estimate volumes, sizes, and projected areas.
        The adopted pole position is within 5$\deg$ of [$\lambda\ \beta$] = [$52\deg -6\deg$]
        or [RA DEC] = [$52\deg\ +12\deg$].}
      {Using two separately determined masses and the volume of our
        hybrid model, we estimate a density of $3.5\pm1.1$ or
        $4.3\pm0.8$ g cm$^{-3}$. From the density evidence
          alone, we argue that this favors an enstatite-chondrite composition, although other compositions
          are formally allowed at the extremes
         (low-porosity CV/CO carbonaceous chondrite or
         high-porosity metallic).  We discuss this in the
           context of other evidence.}

%        either of which favors a link between
%        Lutetia and enstatite- rather than carbonaceous chondrites.}


%\titlerunning{short}
\titlerunning{The triaxial ellipsoid dimensions of (21) Lutetia}

   \maketitle
%
%________________________________________________________________
%%%%%%%%%%%%%%%%%
%%%%%% TAG %%%%%%
%%%%%%%%%%%%%%%%%

\section{Introduction}

  \indent The second target of the ESA Rosetta mission, asteroid (21)
  Lutetia, had a favorable opposition
  in 2008-09, reaching a minimum
  solar phase angle  of $\omega = 0.45\degr$ on 2008 November 30, and a
  minimum distance from the Earth of 1.43 AU a week earlier. Based on previously
  determined sizes of Lutetia, with diameters ranging from 96 km from IRAS
  \citep{2002-AJ-123-Tedesco, PDSSBN-Iras}
  to 116 km from radar \citet{1999-Icarus-140-Magri, 2007-Icarus-186-Magri},
  Lutetia should
  have presented an apparent diameter of 0.10\arcsec, slightly more than twice
  the diffraction limit of the Keck Observatory 10 m telescope at infrared
  wavelengths (1-2 $\mu$m).
  Continuing our campaign
  \citep{2007-Icarus-191-Conrad,
    2009-Icarus-202-Drummond,
    2010-Icarus-205-Carry-a}
  to study asteroids resolved
  with the
  world's large telescopes equipped with adaptive optics (AO), we have
  acquired
  more than 300 images of Lutetia, most from the 2008-09 season.
  An exceptionally good set of 81 images was obtained on 2008 December 2
  with the Keck II telescope, which, despite the high sub-Earth
  latitude, yields a full triaxial ellipsoid solution from the changing
  apparent ellipses projected
  on the plane of the sky
  by the asteroid. Analyzing all available images
  (2000, 2007, and 2008-09 seasons)
  yields a result consistent with the December 2 set.

\section{Observations}
  \indent
  Table~\ref{tab: log}
  gives the observing circumstances for all
  seven observation dates, where, in addition to the date, Right Ascension,
  Declination, and Ecliptic longitudes and latitudes of Lutetia, we give
  its Earth and Sun distance, the solar phase angle ($\omega$), the
  position angle of the Sun measured east from north while looking at
  the asteroid (NTS), the filter and the number of images ($m$) used
  to form mean  measurements at $n$ epochs, the scale in km/pix at
  the distance of the asteroid on the date, and the instrument and telescope
  configurations used for the observations as listed in
  Table~\ref{tab: config}.
  Multiplying the first scale in Table~\ref{tab: log} by the scale in
  Table~\ref{tab: config} will give
  a km/$\arcsec$ scale at the distance of the asteroid, and multiplying
  this by the appropriate resolution element ($\Theta$) from
  Table~\ref{tab: config}, which is the
  diffraction limit $\Theta = \lambda/D$, where $\lambda$ is the wavelength and
  $D$ is the telescope diameter, gives the last column of Table~\ref{tab: log}, the km per resolution element scale.

\begin{figure}%[thb]
  \includegraphics[width=.5\textwidth]{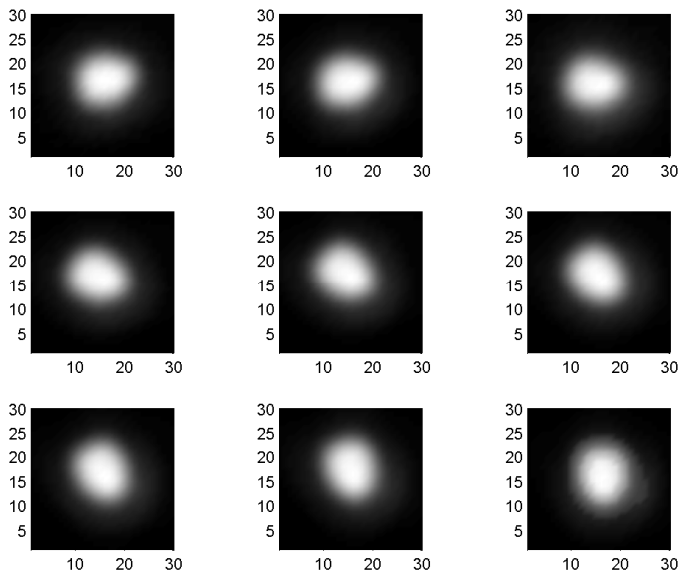}
   \caption{Sample AO images of Lutetia at nine epochs on 2008 December 2. The scale is in pixels, where 1 pixel corresponds to
     0.01\arcsec. North is up and
     east is to the left. Images are displayed on a linear scale, after bi-linear interpolation. From left to right, top to bottom, the images were obtained between 6:52 and 9:01 UT after subtracting for light-time travel. They correspond to the 9 points in Fig ~\ref{fig: dec}.}
   \label{fig: images}
\end{figure}

  \indent The best data set, obtained at K$^\prime$ on December 2, comprises images
  at 9 epochs, 9 images per epoch, where each image is a 0.4 second exposure.
  Figure~\ref{fig: images} shows a single image from each of the 9
  epochs, and clearly
  reveals the asteroid rotating over a quarter of its 8.2 h period.

\begin{figure}
  \includegraphics[width=.5\textwidth]{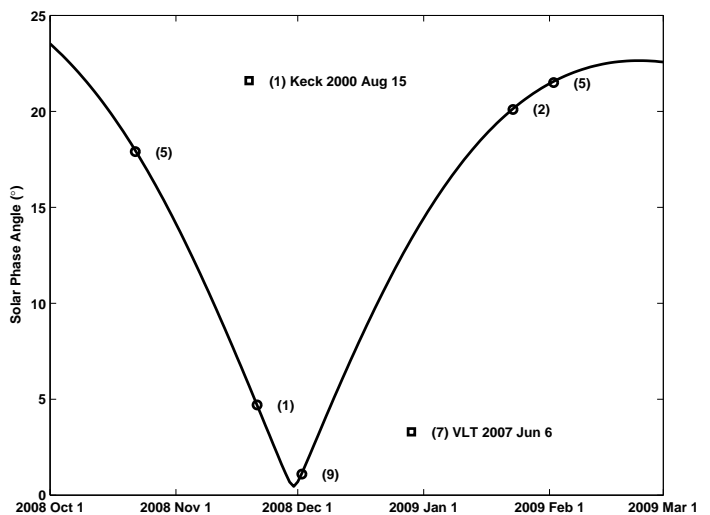}
   \caption{Range of solar phase angles ($\omega$)
     for (21) Lutetia observations.
     All (circles) but
     two dates (squares)
     are from the 2008-09 opposition.
     We list the date and observatory for observations obtained
     outside the 2008-09 campaign.
     The number of epochs for
     each date are listed in parentheses.
   }
   \label{fig: phase}
\end{figure}

   \indent Figure~\ref{fig: phase} illustrates the range of solar phase angles for our
  observations, showing that on more than half of the dates the phase
  angle was greater than $17\degr$, which, for irregular bodies, can
  lead to violations of the assumptions we adopt in
  Section~\ref{sec: analysis}. In this particular case, however, we have found the data are still quite useful and contribute substantially to our final results.

%\begin{landscape}
\begin{table*}
\caption{Observation Log for (21) Lutetia}
\label{tab: log}
\centering
\begin{tabular}{crrccccccccc}
\hline\hline
Date & EQJ2000\ \ \ & ECJ2000\ \ \ & V & Earth & Sun & $\omega$ & NTS &  Filter &  Scale1  &Config&Scale2\\
(UT) & (RA\degr~Dec\degr)& (Lon\degr~Lat\degr) & & (AU)  & (AU) & (\degr)& (\degr)  &   (m/n)   &(km/pix) &Table 2 & (km/rex)\\
\hline
Aug 15, 2000           & 12.3  $-\phantom{0}0.3$ & 11.2 $-5.1$ & 10.5 & 1.239 & 2.057 & 21.6 &  71.6 & K$^\prime$(19/1)    & 15.10 & A &39.8\\
Jun \phantom{2}6, 2007 & 246.0  $-20.7$ & 247.8  $+0.9$        & 10.1 & 1.295 & 2.305 &  3.3 & 286.4 & Ks(35/7)            & 12.47 & B &51.7\\
Oct 22, 2008           & 75.1  $+20.9$ & 76.2  $-1.8$          & 11.2 & 1.555 & 2.352 & 17.9 &  85.9 & J$\phantom{s}$(15/2)          & 11.21   & C &29.3\\
Oct 22, 2008           & 75.1  $+20.9$ & 76.2  $-1.8$          & 11.2 & 1.555 & 2.352 & 17.9 &  85.9 & H$\phantom{s}$(15/2)          & 11.21 & C &38.3\\
Oct 22, 2008           & 75.1  $+20.9$ & 76.2  $-1.8$          & 11.2 & 1.555 & 2.352 & 17.9 &  85.9 & K$^\prime$(33/3) & 11.21 & C &49.6\\
Nov 21, 2008           & 69.4  $+20.8$ & 70.9  $-1.3$          & 10.5 & 1.430 & 2.406 &  4.7 &  88.5 & K$^\prime$($\phantom{5}$4/1)     & 10.31 & C &45.6\\
Dec \phantom{2}2, 2008 & 66.4  $+20.6$ & 68.1  $-1.1$          & 10.2 & 1.441 & 2.426 &  1.1 & 236.7 & K$^\prime$(81/9) & 10.39  & C &46.0\\
Jan 23, 2009           & 59.6  $+20.6$ & 61.9  $+0.1$          & 11.8 & 1.895 & 2.518 & 20.1 & 258.5 & K$^\prime$(30/2)     & 13.66 & C &60.5\\
Feb \phantom{2}2, 2009 & 60.6  $+20.9$ & 62.8  $+0.2$          & 12.0 & 2.033 & 2.534 & 21.5 & 258.9 & H$\phantom{s}$(30/2)          & 14.66  & C &81.1\\
Feb \phantom{2}2, 2009 & 60.6  $+20.9$ & 62.8  $+0.2$          & 12.0 & 2.033 & 2.534 & 21.5 & 258.9 & K$^\prime$(45/3) & 14.66 & C &64.9\\
\hline
\end{tabular}
\end{table*}
%\end{landscape}

\begin{table*}
\caption{Configuration for Table~\ref{tab: log}}
\label{tab: config}
\centering
\begin{tabular}{cccccccc}
\hline\hline
Configuration & Instrument  & Telescope  & Aperture & Scale & Filter  & Wavelength &Resolution \\
& & & (m) &  (pix/\arcsec) & & ($\mu$m) & (\arcsec)\\
\hline
A & NIRSPEC & Keck II     & 10   & 59.52 & K$^\prime$ & 2.12  & 0.044 \\
B & NACO    & ESO VLT UT4 & 8.2  & 75.36 & Ks         & 2.18  & 0.055 \\
C & NIRC2   & Keck II     & 10   & 100.6 & J          & 1.25  & 0.026 \\
  &         &             &      &       & H          & 1.63  & 0.034 \\
  &         &             &      &       & K$^\prime$ & 2.12  & 0.044 \\
\hline
\end{tabular}
\end{table*}

\section{Analysis\label{sec: analysis}}

  \indent Assuming that an asteroid can be modeled as a uniformly illuminated
  triaxial ellipsoid rotating about its short axis, it is possible to
%  convert a series of ellipses projected as it rotates into a full
%  triaxial ellipsoid and find the direction of its spin axis
  estimate its diameters
  ($a \geq b \geq c$)
  and find the direction of its spin axis
  from the observation of a series of ellipses projected as it rotates.  We can
  thus treat any asteroid in the same manner as we treat asteroids that are well
  described by the triaxial assumption, e.g., those that we have defined as
  Standard Triaxial Ellipsoid Asteroids, or STEAs
  \citep{1985-Icarus-61-Drummond,
    1998-Icarus-132-Drummond,
    2009-Icarus-202-Drummond,
    2007-Icarus-191-Conrad,
    2008-Icarus-197-Drummond}.
  The key to turning the projected ellipses into a triaxial
  ellipsoid is determining the ellipse parameters from AO images. We use
   the method of Parametric Blind Deconvolution
  \citep[PBD,][]{1998-Icarus-132-Drummond, 2000-LGSAO-Drummond}
   to find the long ($\alpha$) and short
  ($\beta$) projected (plane of sky) ellipse axes dimensions and the orientation or
  position angle (PA) of the long axis. PBD allows us to find not only the
  asteroid
  ellipse parameters but parameters for the Point Spread Function (PSF)
  as well.
  %(We report their FWHM in Table~\ref{tab: log}).
  Having shown that a good model for the AO PSF is a Lorentzian
  \citep{1998-Icarus-132-Drummond, 2000-LGSAO-Drummond}
  we simply treat a disk-resolved
  image of an asteroid as the convolution of a flat-topped ellipse and a
  Lorentzian, making a simultaneous fit for each in the Fourier plane
  where the convolution becomes a simple product.  \\
  \indent All 307 images were fit for the projected asteroid ellipse parameters
  and individual Lorentzian PSFs. The mean and standard deviations of
  the parameters were formed around epochs consisting of a series of 3 to 15
  images obtained in one sitting at the telescope.
  We then solve the triaxial ellipsoid from a least square
  inversion of the ellipse parameters
%  These parameters were
%  then used in a least squares program that converts the projected
%  ellipses into a triaxial ellipsoid
  \citep{1985-Icarus-61-Drummond, 2000-LGSAO-Drummond}.

  The projected ellipse parameters
  can also be extracted from
  images deconvolved with an alternate algorithm such as
  \textsc{Mistral} \citep{2000-Msn-99-Conan,2004-JOSAA-21-Mugnier}.
  Such contours provide a direct measurement of details of the
  projected shape of the
  asteroid, allowing the construction of the radius vector model that
  we present in \citet{2010-AA--Carry}, providing a more refined
  description of the shape of Lutetia.
  %However, since the ellipse parameters found from the contours either
  %nearly exactly matched the parameters from PBD, or else were %wildly
  %scattered around it
  However, only the PBD parameters were used in deriving
  ellipsoid solutions here. \\
  \indent While observations from one night can produce triaxial ellipsoid results
  \citep{2007-Icarus-191-Conrad,
    2008-Icarus-197-Drummond,
    2009-Icarus-202-Drummond}
  it is possible to combine observations from different nights over multiple
  oppositions to make a global fit if a sidereal period is known
  with enough accuracy
%arc_1  (Drummond et al. 2010).
  (Drummond et al. in preparation).
%arc_1  \tbd{Jack , what are you refering to here??? if it published it
%arc_1    should be done with bibtex, if it something to be sumitted, we
%arc_1    have to claim it as such, not with a 2010}
%% I'm trying here different wording to first sell your global
%% inversion, then we will say bad luck for Lutetia
  This can resolve the natural two-fold ambiguity
  in the location of a rotational pole from a single night of data
  (see Section~\ref{sec: pole}), and
  in some cases reduce the uncertainty in the triaxial ellipsoid
  dimensions if the asteroid is observed over a span of sub-Earth
  latitudes.
  For instance, when observations are restricted to high sub-Earth
  latitudes, even a few images at an equatorial aspect will supply a
  much better view of the $c$ axis than a long series at near polar
  aspects. In other words, different viewing geometries generally
  lead to a better solution. \\
%%%% here is your text
%%  This can resolve the natural two-fold ambiguity
%%  \com{maybe add few words to explain the ambiguity}
%%  in the location of a rotational pole from a single night of data, and
%%  in some cases reduce the uncertainty in the triaxial ellipsoid
%%  dimensions when observations are restricted to high sub-Earth
%%  latitudes, for even a few images at an equatorial aspect will supply a
%%  much better view of the $c$ axis than a long series at near polar
%%  aspects. In other words, different viewing geometries can lead to a
%%  better solution. \\
%
  \indent Unfortunately, during the last two oppositions, in 2007 and 2008-09, Lutetia's positions were $\sim180\degr$ apart on the celestial sphere. In fact, the position of Lutetia for the Very Large Telescope (VLT) observations on 2007 Jun 6 was exactly $180\degr$ from its position on 2008 December 2 (see Table~\ref{tab: log}),
  which meant that regardless of the location of the rotational pole, the two sets of observations were obtained at the same sub-Earth latitude, but of opposite signs. Thus, with the now-known pole, our observations of Lutetia over the 2008-09 opposition were at the same (but southerly) deep sub-Earth latitudes as the VLT observations in the previous (but northerly) deep sub-Earth latitude opposition in 2007, affording the same view of the strongly fore-shortened $c$ axes, but from the opposite hemisphere. The single set of images in 2000 was also obtained at deep southerly latitudes. Thus, we
  have no equatorial view of Lutetia and, therefore, its shortest ($c$) dimension remains less well determined than the other two.

\section{Results}

\subsection{Ellipsoid Fits}
  \indent We made two separate fits to our data, one using only our
  best data set from December 2, and the other combining all of our
  data taken over the 8.5 year span using a sidereal period of
  8.168\,27 h from \citet{2010-AA--Carry} to link them together.
    The results of these two fits give highly consistent values for
    the equatorial dimensions, but not for $c$.
    When fitting all of the data, a higher $c$ value is
    preferred, but we restrict $c$ to the usual physical
    constraint of $b \ge c$.  Because the entire data
    set, taken as an ensemble, has higher noise, we
    consider the $b=c$ as a limiting case (a biaxial
    ellipsoid), and adopt the value of c derived from
    the Dec 2 data alone as our preferred value.
 Both solutions are listed in Table~\ref{tab: stea}, where $\theta$ is the sub-Earth latitude, $PA_{Node}$ is the position angle of the line of nodes measured east from north, and $\psi_0$ is rotational phase zero, the time of maximum projected area when the $a$ axis lies unprojected in the plane of the sky. The uncertainties shown in Table 3 are the internal precisions of the model fit, and do not include systematics. We have assigned overall uncertainties to our
    ellipsoid model, including systematics, of $4 \times 3\times 13$ km in the dimensions, and 5$\degr$ in the pole position.

\begin{table}
\caption{(21) Lutetia Ellipsoid Fit Solutions}
  \label{tab: stea}
\centering
\begin{tabular}{c|c|c}
\hline\hline
& Triaxial (Dec 2008) & Biaxial (All)\\
\hline
%%% my version
a (km) & 132$\pm$1 & 132$\pm1$\\
b (km) & 101$\pm$1 & 101$\pm1$\\
c (km) &  93  $\pm$8 & c=b\\
$\theta$ (\degr)    & $-66\pm$3 & $-59\pm3$\\
$PA_{Node} (\degr)$ & 155$\pm3$ & 178$\pm6$\\
$\psi_0$ (Max) (UT) &5.66$\pm0.07$& 5.10$\pm0.12$\\
\hline
Pole&&\\
\ [RA\degr; Dec\degr] & [44;+9] & [34; +16]\\
$\sigma$ radius (\degr)& 2.5 & 3.1\\
\ [$\lambda\degr; \beta\degr$] &[45;-7] & [37; +3]\\
%
%%%% yours
%a (km) & 132.5$\pm$1 & 132.5$\pm1$\\
%b (km) & 100.5$\pm$1 & 100.5$\pm1$\\
%c (km) &  93  $\pm$8 & = b\\
%$\theta$& $-66\degr\pm$3 & $-59\degr\pm3$\\
%$PA_{Node}$&155$\degr\pm3$ & 178$\degr\pm6$\\
%$\psi_0$ (Max) UT&5.66$\pm0.07$& 5.10$\pm0.12$\\
%\hline
%Pole&&\\
%\ [RA; Dec] & [44$\deg$;$+9\deg$] & [34$\deg$; $+16\deg$]\\
%$\sigma$ radius& 2.5$\deg$ & 3.1$\deg$\\
%\ [$\lambda; \beta$] &[45$\deg$;$-7\deg$] & [37$\deg$; $+3\deg$]\\
\hline
\end{tabular}
\end{table}

%%% I think A&A prefers the table that way
\begin{table}
\caption{RMS of Projected Ellipses from Models}
\label{tab: error}
\centering
\begin{tabular}{c|c|c|c|c}
\hline\hline
 & $\alpha$ & $\beta$ &  PA & Fig\\
 & (km) & (km) & (\degr) \\
\hline
Triaxial (2008 Dec only) & 0.4 & 0.9 & 1.1 & 3 \\
Triaxial (All)      & 4.1 & 4.7 & 6.9 & 4\\
Biaxial (All)       & 3.4 & 4.4 & 6.6 & 5\\
\hline
\end{tabular}
\end{table}

%%% your version
%\begin{table}
%\caption{Lutetia Projected Ellipse Uncertainties}
%\label{tab: error}
%\centering
%\begin{tabular}{c|c|c|c}
%\hline\hline
%& Tri (Dec 2008 only) & Tri (All) &Bi (All) \\
%%&  Dec 2008 only & All & All \\
%\hline
%$\alpha$& 0.4 km& 4.1 km  & 3.4 km\\
%$\beta$ & 0.9 km& 4.7 km  & 4.4 km\\
%PA      & 1.1$\deg$ &6.9$\deg$ &6.6$\deg$\\
%\hline
%\end{tabular}
%\end{table}

\begin{figure}
  \includegraphics[width=.5\textwidth]{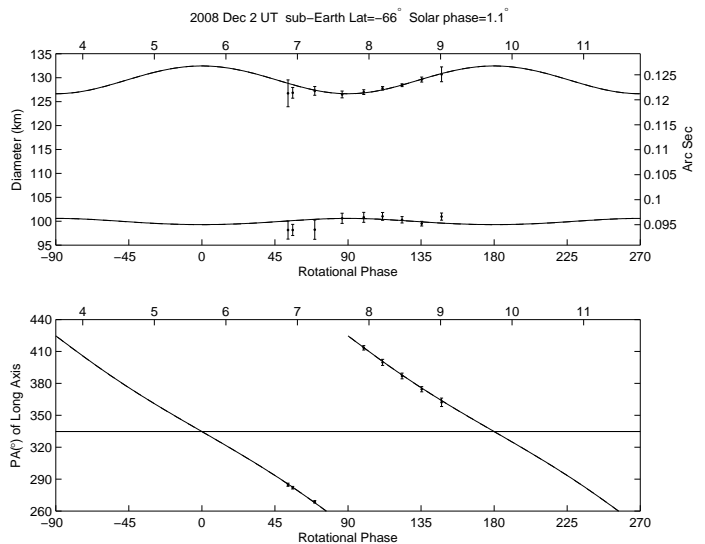}
   \caption{Triaxial ellipsoid fit to
     measured ellipse parameters of (21) Lutetia
     on 2008 Dec 2.
     In the upper subplot, each image's
     long ($\alpha$) and short ($\beta$) axis dimensions are plotted
     as dots with $1\,\sigma$ uncertainties. The
     lines are the prediction for the projected ellipses from the
     triaxial ellipsoid parameters in Table~\ref{tab: stea}, derived from
     the fit to
     the data. Because the solar phase angle
     ($\omega$) is only 1.1\degr~(Fig.~\ref{fig: phase}
     and Table~\ref{tab: log}), the ellipse parameters for the
     terminator ellipse is coincident with the projected ellipse lines. The lower
     subplot shows the same for the position angle of the long axis, where
     the horizontal line is the line of nodes, the intersection of the
     asteroid's equator and the plane of the sky.
     This figure is corrected for
     light time travel, \textsl{i.e.}, the plot is in the body-centered time
     frame.}
\label{fig: dec}
\end{figure}

\begin{figure}
  \includegraphics[width=.5\textwidth]{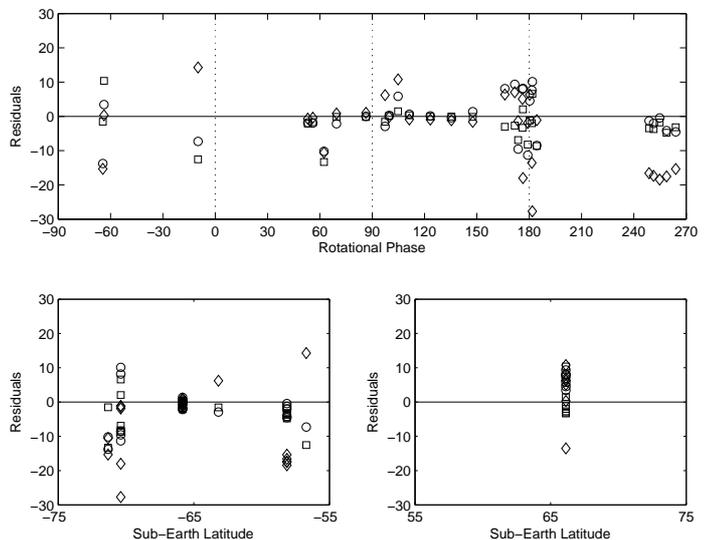}
   \caption{Residuals between the triaxial ellipsoid model (obtained from December 2008 observations only) and all observations
     as function of the rotation phase and latitude.
     The major and minor axes residuals are shown as squares
     and circles, respectively, both in km, and the position angle
     residuals, in degrees, are shown as diamonds. From
     Table~\ref{tab: error}, the rms
     of the weighted residuals are 4.1 and 4.7 km for the major and minor
     axes, respectively, and 6.9$\deg$ for the position angle. The data
     in the northern hemisphere is from the VLT in 2007.}
   \label{fig: stea_vs_all}
\end{figure}

\begin{figure}
  \includegraphics[width=.5\textwidth]{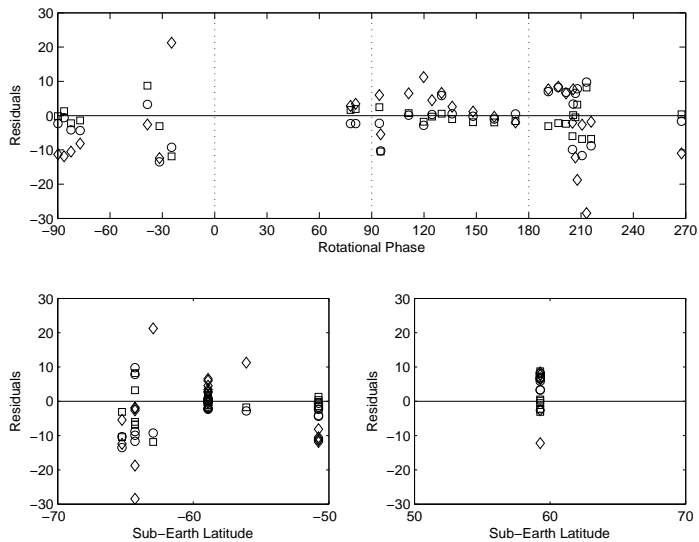}
   \caption{Same as previous figure, except that the residuals are between the biaxial ellipsoid model and the data. From Table~\ref{tab: error},
     the rms of the weighted
     residuals are 3.4 and 4.4 km for the major and minor axes,
     respectively, and 6.6$\deg$ for the position angle.}
   \label{fig: bi_vs_dec}
\end{figure}

Figure~\ref{fig: dec} shows the triaxial ellipsoid fit to the December
data. Figure~\ref{fig: stea_vs_all}
shows the residuals from the prediction using the December triaxial
ellipsoid model for all of the data,
%(The triaxial ellipsoid model is
%plotted over all of the data in the Appendix.)
%For the biaxial
%ellipsoid case, rather than show the projected model on all seven
%dates,
and Figure~\ref{fig: bi_vs_dec} shows
%just
the overall
residuals to the biaxial fit. The difference
between Figs~\ref{fig: stea_vs_all}
and~\ref{fig: bi_vs_dec} is subtle, showing that Lutetia
is close to a      % I prefer it this way
prolate ellipsoid. Trends in some of the residuals, such as the
curling set of position angles at the right of the top plot in
Fig~\ref{fig: stea_vs_all},
and at the left of the similar plot in Fig~\ref{fig: bi_vs_dec},
indicate departures
from our assumptions of a smooth featureless ellipsoid rotating about
its short axis, and motivate
the more detailed shape model that we present
in \citet{2010-AA--Carry}.
The rms weighted (by the observational uncertainty of each measurement)
residuals for $\alpha$ and $\beta$,
the projected ellipse major and minor axes lengths, and the weighted
residuals for the position angle (PA) of the long axis, are given in
Table~\ref{tab: error}, and are to be associated with Figs 3-5.
These can be interpreted as uncertainties (but without possible systematics) for any predicted future
projected ellipse parameters.

The axial ratios derived from our model are $a/b=1.32$ and $b/c=1.09$.  From a
compilation\footnote{\href{http://vesta.astro.amu.edu.pl/Science/Asteroids/}
  {http://vesta.astro.amu.edu.pl/Science/Asteroids/}
  is a web site gathering sidereal periods, rotational poles, and axial
  ratios maintained by A. Kryszczy{\'n}ska. See
  \citet{2007-Icarus-192-Kryszczynska}.} of axial ratios and
rotational poles, mostly from lightcurves,
the average axial ratios are $a/b=1.27\pm0.06 $ and
$b/c=1.45\pm0.55$, both within one sigma
of our directly determined values.
%While our axial ratio of $a/b=1.32\pm0.02$ falls
%within this range, the average $b/c$ ratio is too uncertain to %compare to ours.
More recent work suggests a $b/c$ ratio of less than 1.1
\citep{2010-AA-Belskaya}, also consistent with our value.
Our fit of the radius vector
model derived from lightcurves by
\citet[][see section~\ref{sec: lci}
below]{2003-Icarus-164-Torppa}
yields an $a/b$ ratio of $1.17$ and $b/c=1.18$,
%$a/b$ ratio of $1.17\pm0.01$ and
%$b/c=1.18\pm0.01$.
and the latest radius vector model derived by
\citet{2010-AA--Carry} from a combination of present AO images and
%historical
lightcurves has ratios of 1.23 and 1.26. Our hybrid model, given as the final result of our paper here (see next section),
has $a/b = 1.23$ and $b/c = 1.09$.

%$1.21\pm0.03$
%and $b/c=1.23\pm0.03$.
%However, the latest model
%\citep{2010-AA--Carry}, which
%takes our AO observations into account, has ratios of $1.21\pm0.04$
%and $b/c=1.34\pm0.04$.

%Considering the lack of AO disk-resolved observations anywhere near Lutetia's equator, and the fact that the total sum of data had to be constrained to a biaxial ellipsoid, for some purposes it may be prudent to consider $c$ as undetermined. In fact, the biaxial ellipsoid solution confirms the $a$ and $b$ dimensions of the triaxial solution, but provides no information about the $c$ dimension, and, therefore, the best estimate of the length of the $c$ dimension comes from the triaxial solution derived from 2008 December 2 data only.
%arc_1
%Overall, the triaxial ellipsoid dimension uncertainties in Table~\ref{tab: stea} should perhaps be increased to $4\times 3\times 13$ km in order to include estimated systematic effects.

%\clearpage

\subsection{Rotational Pole\label{sec: pole}}

  \indent There is a natural two-fold ambiguity in the location of the
  rotational pole with our method that is symmetric with respect to the   position of the asteroid if observed on one night. Thus, there are two possible poles for the December triaxial ellipsoid solution. However, if the asteroid can be observed at significantly different positions, then the rotational pole can be disambiguated.
%  Hence, the 2008-09 observations and the 2007 data from the same
%  positions (Table~\ref{tab: log}) do not provide enough diversity to
%  break the ambiguity, but thanks to the single 2000
%  observation, the ambiguity is broken in the biaxial case since the
%  residuals are some 18\% higher
%  for the rejected pole than for the accepted region.
%
%       I prefer the original
  The single 2000 observation helps break the ambiguity since the residuals are some 18\% higher
  for the rejected pole than for the accepted region when considering all of the data. Otherwise,
  the 2008-09 observations and the 2007 data from the same positions,
  would not have provided enough diversity to break the ambiguity.

\begin{figure}
  \includegraphics[width=.5\textwidth]{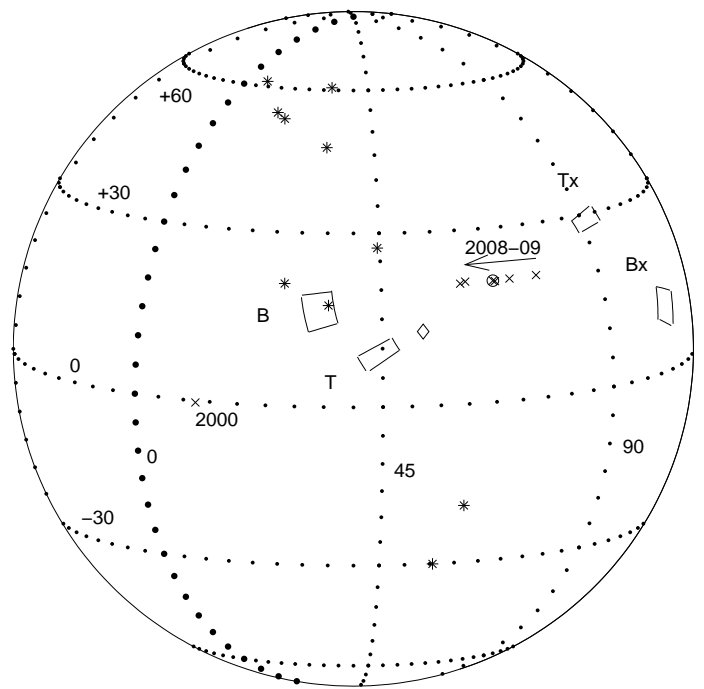}
   \caption{Pole locations for Lutetia on the celestial globe.
     The
     positions of Lutetia on the sky for the seven nights of our observations are
     indicated by $\times$'s. The opposite of the 2007 position is the circle on
     top of the $\times$ for December 2008. The poles found from lightcurve work
     are marked with asterisks, with the LCI pole
     \citep{2003-Icarus-164-Torppa} shown as a
     diamond (our final pole, the KOALA pole \citep{2010-AA--Carry}, is less than a degree from the LCI pole). The four wedge shaped areas are the uncertainty regions
     around our poles, with T and B indicating our two each possible
     triaxial and biaxial ellipsoid solution poles. Our rejected poles
     are marked as Bx and Tx at the far right, while the lightcurve
     rejected pole region is on the other side of the globe.}
   \label{fig: pole}
\end{figure}

  The poles from various lightcurve techniques can have two- or
  four-fold ambiguities
  \citep[see][for a good summary]{1989-AsteroidsII-2-Magnusson},
  which can be broken when paired with our results.
  Figure~\ref{fig: pole} shows the
  positions of about half of the poles (see footnote 1) found from
  lightcurve methods (the other half lie on the opposite hemisphere), as well as ours. We assert that the correct region for the pole location is
  where our poles near $RA=45\degr$ coincide with the span of lightcurve poles in this hemisphere. Furthermore, the lightcurve inversion (LCI; see
  section~\ref{sec: lci})
  pole of
  \citet{2003-Icarus-164-Torppa}
  \footnote{%We use the updated pole solution available at
    \href{http://astro.troja.mff.cuni.cz/projects/asteroids3D/web.php}
  {http://astro.troja.mff.cuni.cz/projects/asteroids3D/web.php}}
  lies at coordinates RA$=52\deg$ and Dec$=+13\deg$, less than $9\deg$
  from our triaxial ellipsoid pole in Table~\ref{tab: stea}.

%\clearpage
\section{Comparison with Lightcurve Inversion Model\label{sec: lci}}

\begin{figure}%[thb]
  \includegraphics[width=.5\textwidth]{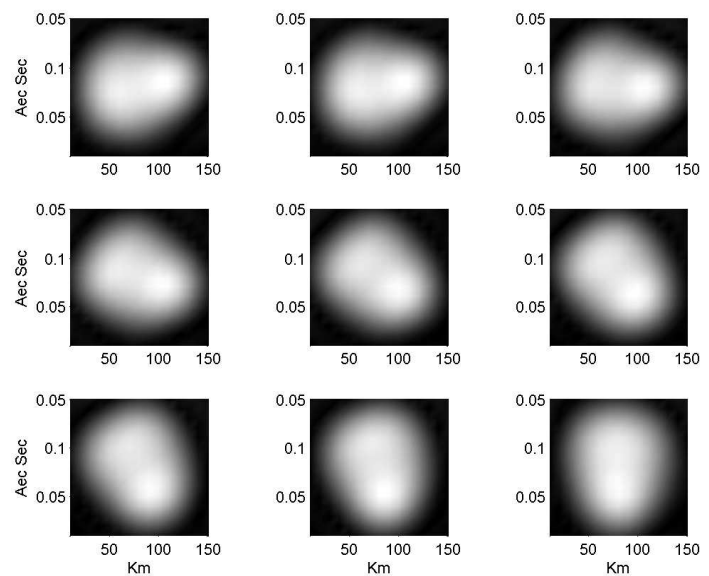}
   \caption{Same as Fig~\ref{fig: images}, except PBD (deconvolved) images of (21) Lutetia from 2008 December 2.}
\label{fig: deconv}
\end{figure}

  \indent Figure~\ref{fig: deconv} shows our PBD images of Lutetia from
  2008 December 2. Each image is the mean of 9 shifted and added images at each
  epoch, and then linearly deconvolved of the Lorentzian PSF found in
  its fit. Notice the tapered end.

\begin{figure}%[thb]
\centering
  \includegraphics[width=.4\textwidth]{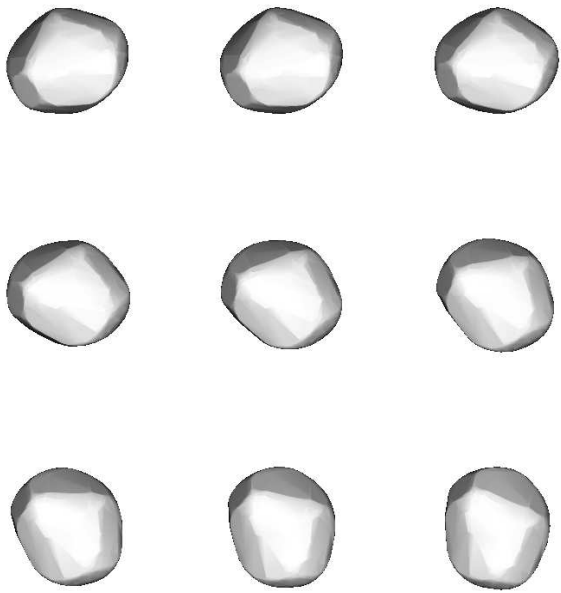}
\caption{Convex shape model of (21) Lutetia from
  \citet{2003-Icarus-164-Torppa}, plotted at same times as
  Fig~\ref{fig: deconv}.}
\label{fig: model}
\end{figure}

  For comparison, the lightcurve inversion model (see footnote 2) based
  on the work of \citet{2003-Icarus-164-Torppa} is shown in
  Fig~\ref{fig: model} for the same
  times. The model appears to match the overall shape and
  orientation in the images, verifying the pole and sidereal period, but it does appear fatter and less tapered than the
  images of the asteroid.
  % underlining the need for an improved shape model.
  In a following article we combine our AO images with lightcurve data
  using a method known as KOALA (Knitted Occultation, Adaptive-optics,
  and Lightcurve Analysis \citep{2010-AA--Carry})
  to produce an improvement over the previous LCI model.
  Not only does it yield better
  matches to the AO images, but it provides an absolute kilometer
  scale, and it can  reproduce Lutetia's lightcurve history.

  Although a triaxial ellipsoid fit of the new KOALA model yields diameters of
  $124\times101\times80$, the model is very non-ellipsoid in appearance,
  and while the $b$ diameter is in agreement between the AO-only and the KOALA model, both the KOALA $a$ and $c$ dimensions are $\sim10$ km smaller than from our triaxial ellipsoid results here.
  The AO-only triaxial ellipsoid solution comes from only a quarter of a rotation, when the minimum area is projected (over what would be a lightcurve minimum). During the time of the 2008 Dec 2 AO observations, the $b$ axis was seen unprojected in the plane of the sky, but both the $a$ and $c$ axes were not. It is the extrapolation, as it were, in rotation to the maximum projected area, when the $a$ axis could be seen unprojected, that leads to an $a$ dimension larger than found from KOALA. The KOALA technique, by combining lightcurves that cover all rotational phases and sub-Earth latitudes with the AO images (at restricted rotational phases and latitudes), finds that there is a large depression on the side of Lutetia away from the 2008 Dec 2 observations that was not completely sampled by our imaging, resulting in the smaller KOALA model $a$ axis dimension.
  This depression of $\sim10$ km explains the difference between the two $a$ dimensions.

  On the other hand, since KOALA only uses amplitudes from lightcurves,
  and since amplitudes are a strong function of $a/b$ but a weak
  function of $b/c$, the KOALA model $c$ dimension is only weakly
  determined when the AO images at high sub-Earth latitudes only reveal
  a strongly foreshortened $c$ axis. (See \citep{2010-AA--Carry} for a
  discussion on the limits of the KOALA inversion in the particular
  case of Lutetia.) Therefore, Lutetia's $c$ dimension is best
  determined from the 2008 December 2 AO data set.

  To make the best possible model for use in evaluating sizes,
  cross-sectional areas, volumes, and densities, we combine what we
  consider the best aspects of both models into a hybrid triaxial
  ellipsoid/KOALA model that has dimensions of $124\times101\times93$
  km, taking the $a$ diameter from KOALA and the $c$ diameter from the
  AO triaxial ellipsoid fit. The original KOALA model radius vector $Z$
  components are merely expanded by 93/80. We estimate the uncertainties
  on these dimensions, including possible systematics, to be
  $5\times4\times13$ km. Our best final average diameter is, then, $(abc)^{1/3}=105\pm5$ km. The rotational pole for these hybrids should be
  the KOALA pole at [RA Dec]=[$52\deg\ +12\deg$], since it is primarily
  based on lightcurves obtained over 47 years as opposed to the triaxial
  ellipsoid pole in Table 3 from nine epochs on 2008 December 2,
  although they are less than 8$\deg$ apart. The uncertainty for this
  pole is about $5\deg$ in each coordinate.

\section{Taxonomy and Density}
  \indent Lutetia was well observed in the 1970s, yielding visible and near-infrared
  reflectance spectra \citep{1975-ApJ-195-McCord},
  radiometric albedos and
  diameter estimates \citep{1977-Icarus-31-Morrison},
  and polarimetric albedos and
  diameter estimates \citep{1976-AJ-81-Zellner},
  which have been confirmed by
  similar observations reported during the last decade
  \citep[see review by][]{2010-AA-Belskaya}.
  Based on these data, \citet{1975-Icarus-25-Chapman}
  placed only three asteroids, (16) Psyche, (21) Lutetia,
  and (22) Kalliope, into a distinct taxonomic
  group to which \citet{1976-AJ-81-Zellner}
  assigned the letter ``M''.
  The M type was
  defined in terms of spectral and albedo properties by
  \citet{1978-Icarus-35-Bowell}, who assigned a diameter of 112 km to
  Lutetia (estimates by
  \citet{1977-Icarus-31-Morrison} and \citet{1976-AJ-81-Zellner} had
  been diameters of 108--109 and 110 km, respectively).\\
%BEN%  These diameters are associated with moderate visual albedos, much higher
%BEN%  than for the more common (CI/CM) carbonaceous chondrites (CC) and
%BEN%  overlapping the lower range for S-types.\\
%
% -- until here it's 100%
% -- but then the subject jump from albedo to meteorites back to ...
% -- it's not linear
%
  \indent It was later found by radar that some, but
  not all, M-types were metallic.
  \citet{1995-Icarus-117-Rivkin}
  recognized that there were two sub-types of M-type
  asteroids. The standard M types showed high radar reflectivity
  and relatively neutral colors, both
  apparently due to metal. The other type (also
  showing similar colors, but now thought to be due to metal flakes
  embedded in a colorless stony matrix) had a
  3 micron band, ascribed to hydrated minerals and which
  was deemed to be unlikely on a chiefly metallic body.
  \citet{1995-Icarus-117-Rivkin} called this new ``wet'' subclass M(W) and
  assigned Lutetia to this subclass \citep{2000-Icarus-145-Rivkin}.
  \citet{1973-Icarus-19-Chapman} first suggested
  that what we now term an M-type spectrum might be associated with
  enstatite chondrites (ECs) and \citet{2000-Icarus-145-Rivkin}
  suggested a hydrated EC as a plausible composition for Lutetia.
  Recently, \citet{2009-Icarus-202-Vernazza} and (partly)
  \citet{2007-AA-470-Nedelcu} showed that ECs are
  a good match for the visible/near-infrared spectra of Lutetia.\\
  \indent The measured visual albedo for Lutetia has typically ranged over
  15--22\%, much higher than for the more common (CI/CM) carbonaceous chondrites (CC) and
  overlapping the lower range for S-types
  (in recent literature, the early dedicated observations of
  Lutetia have been supplanted by reference to five rather inconsistent
  IRAS scans, which imply a still higher albedo and smaller effective
  diameter for Lutetia, well under 100 km, to which we assign less
  significance, especially because they are inconsistent with the mean
  size derived here).
  The radiometry by \citet{2006-AA-447-Mueller},
  reduced using two different thermal models, also yields albedos too
  high for most CC meteorites.
  A recent determination of visual albedo, using
  Hubble Space Telescope observations
  \citep{2009-arXiv-Weaver} and the size/shape/pole
  results from the present paper and
  \citet{2010-AA--Carry}, indicate a
  value near 16\%. This value, consistent with EC albedos (as well
  as metallic), is generally higher than most CCs, although
  some types of CCs, namely CO/CVs, have higher
  albedos, typically about 10\%, with some COs getting as high
  as 15--17\% \citep{2009-Icarus-202-Clark}.\\
  \indent Radar observations of Lutetia
  (\citet{1999-Icarus-140-Magri, 2007-Icarus-186-Magri}, confirmed by
  \citet{2008-Icarus-195-Shepard})
  showed that Lutetia has a moderate radar albedo (0.19--0.24), comfortably
  in the mid-range of ECs, but considerably
  lower than metallic M-types and
  higher than most CCs.
  The uncertainty range in these values overlaps with some CO/CV carbonaceous chondrites compositions
  at the low extreme and with some metallic compositions at the high end.
  Hence, CO/CV carbonaceous chondrites cannot
    be ruled out based on albedo considerations
    alone.\\
  \indent Indeed, numerous researchers in the last few years
  (\citet{2004-AA-425-Lazzarin,
   2009-AA-498-Lazzarin,
   lazzarin-2010,
   2005-AA-430-Barucci,
   2006-AA-454-Birlan,
   2008-AA-477-Barucci,
   2010-AA-513-Perna},
  see summary by \citet{2010-AA-Belskaya}) have argued that Lutetia
  shows certain spectral characteristics
  (\textsl{e.g.}, in the thermal IR) that
  resemble CO and CV types and do not resemble
  a metallic meteorite. However, in these studies, the comparisons with EC
  meteorites was less thorough, partly because mid-infrared comparison data
  are not extensive.
  From rotationally resolved visible/near-infrared spectra of Lutetia,
  \citet{2007-AA-470-Nedelcu} claimed a better match with CC
  in one hemisphere and with EC in the other, although this
      has yet to be confirmed.  If Lutetia were
      highly heterogenous, that might explain some
      of the conflicting measurements.
\\
  \indent \citet{2010-Icarus-207-Vernazza}, however, have shown that
  mid-infrared emission of asteroids of similar composition can be
  very different due to differences in surface particle size.
  Mineralogical interpretations from this wavelength range are thus
  subject to caution and must be supported by VNIR reflectance
  spectra.
  Also, although some CO meteorites show albedos approaching that of
  Lutetia, the lack of a 1 micron olivine band
  in Lutetia's reflectance spectrum
  \citep[see Fig.~3 of][]{2005-AA-430-Barucci}
  argues against CO composition since
  most, but not all, COs have a 1 micron band. Since
  the strength of this band
  generally shows a positive correlation with albedo,
  Lutetia's high albedo suggests that a strong 1 micron band should be present if its composition were CO.
  The lack of a drop-off in Lutetia's
  spectral reflectance below 0.55 micron
  and its high albedo
  make it inconsistent with CV meteorites
  \citep[see][for instance]{1976-JGR-81-Gaffey}.
  Finally, M and W-type asteroids
  (parts of the X class \citep{2009-Icarus-202-DeMeo} if albedo is not
  known) have colors in the visible that are inconsistent with
  C-types.\\
  \indent Colors, spectra, polarization, and albedos
  give us a picture of the relatively thin surface layers of an
  asteroid.  Effects such as space weathering, repeated
  impacts that churn the regolith, recent impacts that
  may locally expose fresh material, particles sizes,
  or even differentiation
  processes may hinder our ability to ascertain the bulk
  composition of an object. Bulk density, on the other
  hand, gives us a picture of the entire asteroid body
  and ought to be a powerful constraint on bulk composition
  (subject to uncertainties about porosity and interior
  structure, mentioned below). Our new size estimates, when combined
  with recent mass determinations from other workers, now
  allow us to make estimates of the bulk density for Lutetia.
  Table~\ref{tab: density} lists the volumes from three of
  the models addressed in this work,
  the triaxial ellipsoid model, the
  KOALA radius vector model, and our
  best estimate hybrid model.
  When coupled with two mass estimates, by \citet{2008-DPS-40-Baer}
  or \citet{2009-AA-507-Fienga}
  (which themselves differ by 25\%), we find the given
  bulk densities.\\
  \indent Grain densities for stony meteorites
  range from $\sim$2.3 g.cm$^{-3}$
  for CI/CM carbonaceous chondrites
  \citep{2008-LPI-39-Consolmagno}
  to $\sim$3.0--3.6 g.cm$^{-3}$
  for CO/CV \citep{1999-Icarus-142-Flynn}
  to $\sim$3.6 g.cm$^{-3}$
  for EC \citep{2009-LPI-40-Macke}.
  Our best model yields densities of 4.3 or 3.5 g.cm$^{-3}$,
  which are among the higher densities yet tabulated
  for asteroids.
  The maximal extent of uncertainties on our preferred
  model range from
  about 2.3 to 5.1 g.cm$^{-3}$.
  Conservatively, if we were to consider meteorite grain
  densities, this range excludes iron-nickel, CI, and CM.
  Enstatite chondrites are favored, but CO/CV  are
  also allowed.
  Perhaps more realistically, meteorite bulk densities
  should be considered instead.  These are significantly
  lower -- \textsl{e.g.}, CO/CV appear to be, on average, about 0.6
  g.cm$^{-3}$ lower
  \citep{2009-LPI-40-Macke}.\\
  \indent One must also be careful
  in trying to put too much emphasis on
  comparison of asteroid
  densities with meteorite densities.
  Most asteroids are thought to have
  significant macroporosity
  \citep[\textsl{e.g.}, they
  may be rubble piles like (25143)
  Itokawa][]{2006-Science-312-Fujiwara}, so the asteroid
  density is likely to be substantially
  lower than the component material
  density
  \citep[see][]{2006-LPI-37-Britt}.
  If so, then metallic meteorites are still ruled out,
  but only marginally at the upper end of the uncertainty
  range. CO/CV are similarly ruled out at the lower end
  of the uncertainty range, but not by much.
  The meteorite density
  values can only be guidelines.
  This provides some
  constraint on the possible bulk
  composition, but without reliable, smaller
  uncertainties in mass estimates, we must also
  rely on other observed quantities,
  such as albedo and spectra.  A
  better mass estimate from Rosetta
  will reduce the density uncertainty
  considerably.\\
  \indent In summary, our consideration of density
  and other evidence favors EC composition
  for Lutetia, and although CO/CV composition
  is not ruled out definitively, we
  consider it a lower probability.
  Among all the known meteorite classes, hydrated enstatite
  chondrites seem to fit the most number of measured
  parameters. These chondrites are represented among
  known meteorites only by the hydrated EC clasts in the
  unusual meteorite Kaidun.
  Finally, we emphasize that Lutetia may well be
  composed of material that is
  either rare or not yet represented in our meteorite collections.
  One example that might work is a low-albedo carbonaceous matrix
  material to suppress the olivine bands, embedded with
  abundant high-albedo clasts (such as an Allende-like composition,
  but with a much higher abundance of CAIs).

\begin{table}
\caption{Lutetia Mass, Volume and Density}
\label{tab: density}
\centering
\begin{tabular}{c|c|c}
\hline\hline
%& \hfill Mass&\\
%& Baer et al. &Mass& Fienga et al.\\
&Vol & Density  (g cm$^{-3}$)\\
%$2.57\pm0.24$ &($\times 10^{21}$ g)& $2.06\pm0.60$\vspace{2pt}\\
Method &($\times 10^{20}$ cm$^3$) & $\rho^a$ \hspace{30pt} $\rho^b$\\
\hline
Triax    & $6.46\pm0.95$ & $3.98\pm0.69$\ \  $3.19\pm1.04$ \\
KOALA & $5.13\pm1.02$ & $5.00\pm1.10$\ \  $4.01\pm1.41$\\
Hybrid& $5.94\pm0.90$ & $4.32\pm0.77$\ \  $3.46\pm1.13$\\
\hline
\end{tabular}
\begin{tabular}{l}
$^a$ with mass of $2.57\pm0.24 \times 10^{21}$ g from
\citet{2008-DPS-40-Baer}\\
$^b$ with mass of $2.06\pm0.60 \times 10^{21}$ g from
\citet{2009-AA-507-Fienga}\\
\end{tabular}
\end{table}

\section{Summary}
  \indent

           We used adaptive optics images of
             (21) Lutetia from various
          large telescope facilities, and at
          various epochs, to make a triaxial
           ellipsoid shape model.  In a
            companion paper, we combine these
             AO images with lightcurves
           covering several decades to produce
          a radius vector model.  There are
           advantages and disadvantages to
            these two methods.  Here, we have
              combined the best properties
           of each to yield a hybrid shape
          model, approximated by an ellipsoid
         of size $124\times 101\times 93$ km (with
         uncertainties $5\times 4\times 13$ km) that
            can be easily used to compute
          sizes, volumes, projected areas,
           and densities.  When coupled
            with recent mass estimates,
           this hybrid model suggests
           a density of $3.5\pm1.1$ g cm$^{-3}$
            or $4.3\pm0.8$ g cm$^{-3}$.  This
           is within the range expected
             for EC-like compositions,
             although the uncertainties
             formally permit other
             compositions.

  %Applying Parametric Blind Deconvolution (PBD) to
  %AO disk-resolved images of some of Saturn's satellites suggests that
  %diameters from PBD
  %should perhaps be increased by $1\pm1$\%, but
  The Rosetta mission presents a unique opportunity for us to perform
  the ultimate calibration of our PBD and triaxial ellipsoid
  approach to determine sizes and rotational poles.
  Furthermore, it will offer a chance to compare and contrast our triaxial ellipsoid model to the KOALA model for Lutetia.
  %Go Rosetta.

\section*{Acknowledgments}
We thank E. Dotta, D. Perna, and S. Fornasier for obtain-
ing and sharing their VLT data, and for spirited discussions concerning Lutetia's taxonomy which we feel materially improved the content of this paper. This study was supported, in part,
by the NASA Planetary Astronomy and NSF
Planetary Astronomy Programs (Merline PI), and
used the services provided by the JPL/NASA Horizons web
site, as well as NASA's Astrophysics Data System.
We are grateful for telescope time made
available to us by S. Kulkarni and M. Busch
(Cal Tech) for a portion of this dataset.  We
also thank our collaborators on Team Keck,
the Keck science staff, for making possible
some of these observations.
In addition, the authors wish to
  recognize and acknowledge the very significant cultural role and
  reverence that the summit of Mauna Kea has always had within the
  indigenous Hawaiian community.  We are most fortunate to have the
  opportunity to conduct observations from this mountain.

%%%--- Bibliography ---%%%
%\bibliographystyle{aa2}
\bibliographystyle{aa}
\bibliography{biblio}

\end{document}